
\magnification=1200
\baselineskip=18pt
\def\ao{\~ao\ }

\def\[{\c c}
\def\\{\'\i }
\def\sqr#1#2{{\vcenter{\hrule height.#2pt
     \hbox{\vrule width.#2pt height#1pt \kern#1pt
      \vrule width.#2pt} \hrule height.#2pt}}}

\font\titlea=cmb10 scaled\magstep1

\font\titlec=cmb10 scaled\magstep3

\hfill IF-UFRJ-21/94
\bigskip
\centerline{\titlec Constrained analysis}
\centerline{\titlec of topologically massive gravity}
\vskip 2.5 cm
\centerline{J. Barcelos-Neto$\,^\ast$ and T.G. Dargam}
\bigskip
\centerline{\it Instituto de F\\sica}
\centerline{\it Universidade Federal do Rio de Janeiro}
\centerline{\it RJ 21945-970 - Caixa Postal 68528 - Brazil}
\vskip 1.5 cm
\centerline{\titlea Abstract}
\bigskip
We quantize the Einstein gravity in the formalism of weak
gravitational fields by using the constrained Hamiltonian method.
Special emphasis is given to the 2+1 spacetime dimensional case where
a (topological) Chern-Simons term is added to the Lagrangian.

\vfill
\bigskip
\hbox to 3.5 cm {\hrulefill}\par
\item{($\ast$)} Electronic mails: ift03001 @ ufrj and barcelos @
vms1.nce.ufrj.br
\eject

{\titlea I. Introduction}
\bigskip
One of the ultimate goals of the quantum field theory (QFT) is the
quantization of gravity. It is well-known that usual rules of
quantization when applied to the Einstein Lagrangian for the
gravitational field do not lead to a finite theory in a sense that
infinities cannot be eliminated by means of regularization and
renormalization procedures~[1]. With the advent of supersymmetry and
strings~[2] there was the expectation of solving all the problems
that the usual QFT was not able to do. Unfortunately, this does not
appear to be the case, unless till now. We already know that
supersymmetry when applied to gravitation (supergravity) does not
lead to a finite theory. On the other hand, the interesting advent of
the strings is nowadays fighting with the complexibility of its
mathematical structures and its second quantization is still far to
be applied in order to obtain reasonable predictions.

\medskip
It has been a common procedure in QFT to go to lower dimensions when
we are trying to understand something in the usual four spacetime
dimensions and we are not succeed. In the case of the Einstein
gravity this procedure usually runs into problems. For example, going
after the trivial one-dimensional case, we have that the Ricci tensor
is identically zero in two-dimensional spacetime. This means that
there is no way of including matter field in a Einstein
bi-dimensional gravity. The three-dimensional case is also
problematic. We find that the final theory does not have any physical
degree of freedom. However, there is an important aspect in this last
case. If one includes a Chern-Simons (CS) term, the final
(topological) theory has one physical degree of freedom and the above
mentioned inconsistency does not exist any more~[3,4].

\medskip
The interesting constraint structure of the Einstein plus CS theory
in 2+1 dimensions is the main motivation of the present work. We
shall consider the approximation of weak gravitational fields~[4]. In
order to better illustrate the general formalism, we consider this
approximation in Sec. II for any space-time dimensions without the CS
term. In Sec. III we add the CS term and restrict ourselves to the
particular case of 2+1 dimensions. Since there are many first-class
constraints in both cases, we have to introduce appropriate
gauge-fixing conditions.  We left Sec. IV for some concluding
remarks.

\vfill\eject
{\titlea II. Hilbert-Einstein theory in a weak gravitational field}
\bigskip
The Hilbert-Einstein Lagrangian reads

$${\cal L}={1\over2\kappa^2}\,\sqrt{-g}\,\,R\,,\eqno(2.1)$$

\bigskip
\noindent where $\kappa$ is the Einstein constant and we shall use
the following flat metric convention ${\rm diag.}\,\,(+,-,\dots,-)$.
At it was said in the introduction, we are not making here any
restriction to the spacetime dimensions $D$.

\medskip
Considering weak gravitational fields we take the expansion around
the flat space geometry as

$$g^{\mu\nu}(x)=\eta^{\mu\nu}
-\kappa\,h^{\mu\nu}(x)\,.\eqno(2.2)$$

\bigskip
\noindent Introducing (2.2) into (2.1) and just keeping free fields
we get

$${\cal L}={1\over4}\,\partial_\lambda h_{\mu\nu}\,
\partial^\lambda h^{\mu\nu}
-{1\over4}\,\partial_\lambda h^\mu_\mu\,
\partial^\lambda h^\nu_\nu
+{1\over2}\,\partial_\lambda h^\lambda_\mu\,
\partial^\mu h^\nu_\nu
-{1\over2}\,\partial_\lambda h^\lambda_\mu\,
\partial_\nu h^{\nu\mu}\,,\eqno(2.3)$$

\bigskip
\noindent where it was considered that the metric is symmetric.
Expression (2.3) is known as the Fierz-Pauli Lagrangian for massless
particles of spin 2. It is a gauge theory and the corresponding gauge
transformation reads

$$\delta h^{\mu\nu}=\partial^\mu\zeta^\nu
+\partial^\nu\zeta^\mu\,.\eqno(2.4)$$

\bigskip
We are going to study the resulting theory described by the
Lagrangian density (2.3) making use of the constrained Hamiltonian
procedure due to Dirac~[5]. Since this is a noncovariant method, it
might be convenient to separate time and space components in the
expression above. We write down the result as

$${\cal L}={1\over4}\,\bigl(\dot h^{ij}\dot h_{ij}
-\dot h^i_i\dot h^j_j\bigr)
+\bigl(\partial^i h^j_j-\partial_j h^{ij}\bigr)\,
\dot h^0_i-V\,,\eqno(2.5)$$

\bigskip
\noindent where

$$\eqalignno{V=&-{1\over2}\,\partial_ih_{0j}\partial^ih^{0j}
-{1\over4}\,\partial_ih_{jk}\partial^ih^{jk}
+{1\over2}\,\partial_ih^{00}\partial^ih^j_j
+{1\over4}\,\partial_ih^j_j\partial^ih^k_k\cr
&-{1\over2}\,\partial_ih^i_j\partial^jh^{00}
-{1\over2}\,\partial_ih^i_j\partial^jh^k\,_k
+{1\over2}\,\partial_ih^{i0}\partial_jh^{j0}
+{1\over2}\,\partial_ih_{ij}\partial^kh^{kj}\,.
&(2.6)\cr}$$

\bigskip
\noindent The canonical momenta are

$$\eqalignno{\pi^{00}
&={\partial{\cal L}\over\partial\dot h_{00}}=0\,,
&(2.7a)\cr
\pi^{0i}&={\partial{\cal L}\over\partial\dot h_{0i}}
={1\over2}\,\bigl(\partial^ih^j_j-\partial_jh^{ij}\bigr)\,,
&(2.7b)\cr
\pi^{ij}&={\partial{\cal L}\over\partial\dot h_{ij}}
={1\over2}\,\bigl(\dot h^{ij}-\eta^{ij}\dot h^k_k\bigr)\,.
&(2.7c)\cr}$$

\bigskip
\noindent There are $D$ primary (first-class) constraints~[5]

$$\eqalignno{\Omega&=\pi^{00}\approx0\,,&(2.8a)\cr
\Omega^i&=\pi^{0i}+{1\over2}\,
\bigl(\partial_jh^{ij}-\partial^ih^j_j\bigr)\approx0\,.
&(2.8b)\cr}$$

\bigskip
\noindent This is in agreement with Castellani assumption that
stablishes that the number of symmetries of the theory (characterized
by each one of the components of $\zeta_\mu$) is the same as the
number of primary first-class constraints~[6].

\medskip
The next step is to look for secondary constraints. We first
construct the primary Hamiltonian density

$$\eqalignno{{\cal H}&=\pi^{00}\dot h_{00}
+2\pi^{0i}\dot h_{0i}+\pi^{ij}\dot h_{ij}
-{\cal L}+\lambda\Omega+\lambda_i\Omega^i\,,\cr
&=\pi_{ij}\dot h^{ij}-{1\over4}\,\dot h_{ij}\dot h^{ij}
+{1\over4}\,h^i\,_i\,\dot h^j\,_j
+\bigl(\lambda+\dot h_{00}\bigr)\,\Omega
+\bigl(\lambda_i+2\dot h_{0i}\bigr)\,\Omega^i+V\,.
&(2.9)\cr}$$

\bigskip
\noindent We notice that is possible to redefine the Lagrange
multipliers $\lambda$ and $\lambda_i$ in order to absorb the
velocities $\dot h_{00}$ and $\dot h_{0i}$. The remaining velocities
can be eliminated by using the momentum expressions (2.7c). The final
result reads

$${\cal H}=\pi^{ij}\pi_{ij}+{1\over2\!-\!D}\,\pi^i_i\,\pi^j_j
+\lambda\Omega+\lambda_i\Omega^i+V\,.\eqno(2.10)$$

\bigskip
\noindent Considering the fundamental Poisson brackets
\footnote{(*)}{It will always be understood that brackets are taken
at the same time $x_0=y_0$.}

$$\eqalignno{\bigl\{h^{00}(x),\,\pi_{00}(y)\bigr\}
&=\delta^{(D-1)}\bigl(\vec x-\vec y\,\bigr)\,,
&(2.11a)\cr
\bigl\{h^{0i}(x),\,\pi_{0j}(y)\bigr\}
&=\delta^i_j\,\delta^{(D-1)}\bigl(\vec x-\vec y\,\bigr)\,,
&(2.11b)\cr
\bigl\{h^{ij}(x),\,\pi_{kl}(y)\bigr\}
&={1\over2}\bigl(\delta^i_k\delta^j_l
+\delta^i_l\delta^j_k\,\bigr)\,
\delta^{(D-1)}\bigl(\vec x-\vec y\,\bigr)\,.&(2.11c)\cr}$$

\bigskip
\noindent and using the Hamiltonian (2.10) we obtain from the
consistency condition the following secondary-constraints

$$\eqalignno{\Phi&=\nabla^2h^i_i
+\partial_i\partial_j\,h^{ij}\approx0\,,&(2.12a)\cr
\Phi^i&=2\partial_j\pi^{ij}
+\partial^i\partial_j\,h^{0i}
+\nabla^2h^{0i}\approx0\,.&(2.12b)\cr}$$

\bigskip
\noindent Introducing these constraints into the Lagrangian by means
of new Lagrange multipliers and using the consistency condition again
we verify that there are no tertiary constraints. We notice that
constraints $\Phi$ and $\Phi^i$ are also first-class.

\medskip
In order to calculate the Dirac brackets we have to fix the gauge.
Let us start from the corresponding of the Coulomb gauge of the
electromagnetic Maxwell theory, i.e.

$$\partial_ih^{ij}\approx0\,.\eqno(2.13)$$

\eject
\noindent We also choose that

$$\eqalignno{h^{00}&\approx0\,,&(2.14a)\cr
h^{0i}&\approx0\,.&(2.14b)\cr}$$

\bigskip
\noindent With the gauge condition $h^{0i}\approx0$, the secondary
constraints $\Phi^i$ just turns to be $\partial_j\pi^{ij}\approx0$.
Compatibility with the equation of motion permit us to infer that

$$\pi^i\,_i\approx0\,.\eqno(2.15)$$

\bigskip
\noindent Summarizing, the full set of (second-class) constraints is

$$\eqalignno{\Omega&=\pi^{00}\approx0\,,&(2.16a)\cr
\Psi&=h^{00}\approx0\,,&(2.16b)\cr
\Omega^i&=\pi^{0i}\approx0\,,&(2.16c)\cr
\Psi^i&=h^{0i}\approx0\,,&(2.16d)\cr
\Phi&=h^i_i\approx0\,,&(2.16e)\cr
\Sigma&=\pi^i_i\approx0\,,&(2.16f)\cr
\Sigma^i&=\partial_jh^{ji}\approx0\,,&(2.16g)\cr
\Phi^i&=\partial_j\pi^{ji}\approx0\,.&(2.16h)\cr}$$

\bigskip
\noindent  This set contains $4D$ constraints. There is no sense
cases with $D\leq2$ because the number of degrees of freedom will be
negative. $D\!=\!3$ also corresponds to a nonphysical case where the
number of degrees of freedom is zero. In general, the number of
degrees of freedom is $D(D\!-\!3)/2$.

\medskip
The calculation of the Dirac brackets is directly done by means of a
hard algebraic work. The nonvanishing ones are

$$\eqalignno{\bigl\{h^{ij}(x),\,
\pi_{kl}(y)\bigr\}=\biggl[{1\over2}
\bigl(\delta^i_l\,\partial^j\partial_k
&+\delta^i_k\,\partial^j\partial_l
+\delta^j_l\,\partial^i\partial_k
+\delta^j_k\,\partial^i\partial_l\bigr)\,
{1\over\nabla^2}\cr
&+{1\over2}\bigl(\delta^i_k\,\delta^j_l
+\delta^i_l\,\delta^j_k\bigr)
-{1\over D-2}\eta^{ij}\eta_{kl}\cr
&-{1\over D-2}\bigl(\eta^{ij}\partial_k\partial_l
+\eta_{kl}\partial^i\partial^j\bigr)\,{1\over\nabla^2}\cr
&+{D-3\over D-2}\,{\partial^i\partial^j\partial_k\partial_l
\over\nabla^4}\biggr]\,\delta^{(D-1)}(\vec x-\vec y\,)\,.
&(2.17)\cr}$$

\bigskip
\noindent We clearly notice that, in fact, the case with $D\!=\!2$
does not make sense by virtue of terms with $(D\!\!-2)^{-1}$.
Further, one can show that for $D\!=\!3$ the above bracket is zero.

\medskip
Looking at (2.17), we observe that there are no problem with
ordering operators, so it can be directly transformed into commutator
by means of the usual rule of quantization

$$\big\{{\rm Dirac\,brackets}\bigr\}\,
\longrightarrow\,{1\over i\hbar}\,
\bigr[{\rm commutators}\bigr]\,.\eqno(2.18)$$

\bigskip
\noindent Considering the transformations above, one can calculate
the propagators among the $h^{ij}$ fields. The result is

$$\eqalignno{<0\vert T\bigl(h^{ij}(-k)\,
h_{kl}(k)\vert0>={i\over k^2}\,\biggl[
\delta^i_k\delta^j_l
&+\delta^i_l\delta^j_k
+{1\over\vec k^2}\Bigl(\delta^i_k\,k^jk_l
+\delta^i_l\,k^jk_k+\delta^j_k\,k^ik_l
+\delta^j_l\,k^ik_k\Bigr)\cr
&-{2\over D-2}\,\Bigl(\eta^{ij}\eta_{kl}
+\eta^{ij}{k_kk_l\over\vec k^2}
-\eta_{kl}{k^ik^j\over\vec k^2}\Bigr)\cr
&+{2(D-3)\over(D-2)}\,{k^ik^jk_kk_l\over\vec k^4}\biggr]\,.
&(2.19)\cr}$$

\bigskip
\noindent We notice that there is no dynamical massive pole.

\vskip1cm
{\titlea III. Topologically massive gravitation}
\bigskip
When the spacetime dimension is odd, it is possible to introduce a CS
term in the Lagrangian. In this section we concentrate on the 2+1
case. The corresponding CS Lagrangian reads

$${\cal L}_{CS}= {1\over\mu}\,\epsilon^{\lambda\mu\nu}\,
\Gamma^\rho_{\lambda\sigma}\,
\bigl(\partial_\mu\,\Gamma^\sigma_{\rho\nu}
+{2\over3}\,\Gamma^\sigma_{\mu\xi}\,
\Gamma^\xi_{\nu\rho}\bigr)\,.\eqno(3.1)$$

\bigskip
\noindent Considering the expansion of weak gravitational fields
given by (2.2) we get

$${\cal L}_{CS}={\kappa^2\over2\mu}\,\epsilon^{\lambda\mu\nu}\,
\Bigl(\partial_\sigma h^\rho_\lambda\,
\partial_\rho\partial_\mu h^\sigma_\nu
-\partial_\sigma h^\rho_\lambda\,
\partial^\sigma\partial_\mu h_{\rho\nu}\Bigr)\,.\eqno(3.2)$$

\bigskip
\noindent Making the separation of space and time components and
adding the result to (2.5) the result is

$$\eqalignno{{\cal L}={1\over4}\,
\bigl(\dot h_{ij}\,\dot h^{ij}
&-\dot h^i_i\,\dot h^j_j\bigr)
+\bigl(\partial^ih^j_j-\partial_jh^{ij}\bigr)\,\dot h^0_i
+{\kappa^2\over\mu}\,\epsilon^{ij}\,
\Bigl(\dot h^k_0\,\partial_k\partial_i\,h_{0j}\cr
&-\partial_i\partial_k\,h_{00}\dot h^k_j
-\partial_ih^k_0\,\ddot h_{kj}
+\ddot h^k_i\,\partial_kh_{0j}
+{1\over2}\,\partial_k\partial_lh^l_i\,\dot h^k_j\cr
&+{1\over2}\,\dot h^k_i\,\ddot h_{kj}
+{1\over2}\,\nabla^2h^0_i\,\dot h_{0j}
+{1\over2}\,\nabla^2h^k\,_i\,\dot h_{kj}\Bigr)-V\,,
&(3.3)\cr}$$

\bigskip
\noindent where

$$\eqalignno{V=&{1\over2}\,\partial_ih_{0j}\partial^ih^{0j}
+{1\over4}\,\partial_ih_{jk}\partial^ih^{jk}
-{1\over2}\,\partial_ih^{00}\partial^ih^j_j
-{1\over4}\,\partial_ih^j_j\partial^ih^k_k\cr
&+{1\over2}\,\partial_ih^i_j\partial^jh^{00}
+{1\over2}\,\partial_ih^i_j\partial^jh^k_k
-{1\over2}\,\partial_ih^i_0\partial_jh^{j0}
-{1\over2}\,\partial_ih^i_j\partial_kh^{kj}\cr
&-{\kappa^2\over\mu}\,\epsilon^{ij}\,\Bigl(
\partial_kh^l_0\partial_i\partial_lh^k_j
-\nabla^2h^k_0\partial_ih_{kj}
-\nabla^2h_{00}\partial_ih_{0j}\Bigr)\,.
&(3.4)\cr}$$

\bigskip
We observe that the inclusion of the CS Lagrangian leads to the
appearance of higher derivative terms~[7]. The correct Hamiltonian
treatment requires a phase space as $h^{\mu\nu}\oplus\dot
h^{\mu\nu}\oplus\pi^{\mu\nu}\oplus s^{\mu\nu}$, where $\pi^{\mu\nu}$
and $s^{\mu\nu}$ are the canonical momenta conjugate to $h_{\mu\nu}$
and $\dot h_{\mu\nu}$, respectively. Now, velocities have to be
considered as independent coordinates. Using the general
expressions

$$\eqalignno{\pi^{\mu\nu}
&={\partial{\cal L}\over\partial\dot h_{\mu\nu}}
-{\partial\over\partial t}\,
{\partial{\cal L}\over\partial\ddot h_{\mu\nu}}
-2\partial_i\,
{\partial{\cal L}\over\partial\,(\partial_i\dot h_{\mu\nu})}\,,
&(3.5)\cr
s^{\mu\nu}&={\partial{\cal L}\over\partial\ddot h_{\mu\nu}}\,,
&(3.6)\cr}$$

\bigskip
\noindent we get

$$\eqalignno{\pi^{00}&=0\,,&(3.7a)\cr
\pi^{0i}&={1\over2}\,\Bigl(\partial^ih^j_j
-\partial_jh^{ji}\Bigr)
+{\kappa^2\over2\mu}\,
\Bigl(2\epsilon^{jk}\partial_j\partial^ih^0_k
-\epsilon^{ij}\nabla^2h_{0j}\Bigr)\,,&(3.7b)\cr
\pi^{ij}&={1\over2}\,\Bigl(\dot h^{ij}
-\eta^{ij}\dot h^k_k\Bigr)
+{\kappa^2\over4\mu}\,\epsilon^{ik}\,\Bigl(
2\partial_k\partial^jh^{00}-\partial^j\partial_lh^l_k
+2\ddot h^j_k\cr
&\phantom{={1\over4}\,\Bigl(\dot h^{ij}\,\,}
-\nabla^2h^j_k
-2\partial_k\dot h^j_0-2\partial^j\dot h^0_k
+(i\,\leftrightarrow\,j)\Bigr)\,,&(3.7c)\cr
s^{00}&=0\,,&(3.7d)\cr
s^{0i}&=0\,,&(3.7e)\cr
s^{ij}&={\kappa^2\over2\mu}\Bigl[\epsilon^{ik}\,
\bigl(\partial_kh^j_0+\partial^jh_{0k}
-{1\over2}\dot h^j_k\bigr)+(i\,\leftrightarrow\,j)
\Bigr]\,.&(3.7f)\cr}$$

\bigskip
\noindent These lead to the following set of primary constraints

$$\eqalignno{\Omega&=\pi^{00}\approx0\,,&(3.8a)\cr
\Omega^i&=\pi^{0i}+{1\over2}\Bigl(
\partial_jh^{ji}-\partial^ih^j_j\Bigr)
-{\kappa^2\over2\mu}\,\Bigl(
2\epsilon^{jk}\partial_j\partial^ih^0_k
-\epsilon^{ij}\nabla^2h^0_j\Bigr)\approx0\,,&(3.8b)\cr
\Lambda&=\pi^i_i+{1\over2}\,\dot h^i_i
-{\kappa^2\over2\mu}\,\epsilon^{ij}\partial_j\partial_kh^k_i
\approx0\,,&(3.8c)\cr
\Theta&=s^{00}\approx0\,,&(3.8d)\cr
\Theta^i&=s^{0i}\approx0\,,&(3.8e)\cr
\Theta^{ij}&=s^{ij}-\Bigl[{\kappa^2\over2\mu}\,\epsilon^{ik}\,
\Bigl(\partial_kh^j_0+\partial^jh_{0k}
-{1\over2}\,\dot h^j_k\Bigr)
+(i\,\leftrightarrow\,j)\Bigr]\,.&(3.8f)\cr}$$

\bigskip
\noindent The total primary Hamiltonian density in this case reads

$${\cal H}=\pi^{\mu\nu}\dot h_{\mu\nu}
+s^{\mu\nu}\ddot h_{\mu\nu}-{\cal L}
+\lambda\Omega+\lambda_i\Omega^i
+\rho\Theta+\rho_i\Theta^i+\rho_{ij}\Theta^{ij}
+\xi\Lambda\,.\eqno(3.9)$$

\bigskip
\noindent Developing the expression above, one can show that some
terms can be absorbed by redefining the Lagrange multipliers. One
interesting point, that might be opportune to be mentioned, is that
$\ddot h_{ij}$ are also eliminated in this way and not by using the
expression (3.7c) that gives $\pi^{ij}$ in terms of $\ddot h^{ij}$
(this expression is not a constraint).  This is providential because
it is not possible to use (3.7c) in order to express $\ddot h^{ij}$
in terms of $\pi^{ij}$ and other components of the phase space. This
is so because the coefficient of $\ddot h_{ij}$ does not have
inverse.

\medskip
The final expression for the primary hamiltonian density reads

$$\eqalignno{{\cal H}=\pi^{ij}\dot h_{ij}
-{1\over4}\,\Bigl(\dot h^{ij}\dot h_{ij}
&+2\partial_ih_{0j}\partial^ih^{0j}
+\partial_ih_{jk}\partial^ih^{jk}
-\dot h^i_i\dot h^j_j\cr
&-2\partial_ih^{00}\partial^ih^j_j
-\partial_ih^j_j\partial^ih^k_k
+2\partial_ih^i_j\partial^jh^{00}\cr
&+2\partial_ih^i_j\partial^jh^k_k
-2\partial_ih^i_0\partial_jh^{j0}
-2\partial_ih^i_j\partial_kh^{kj}\Bigr)\cr
+{\kappa^2\over2\mu}\,\epsilon^{ij}\Bigl(
2\partial&_i\partial_kh^{00}\dot h^k_j
-2\partial_kh^l_0\partial_l\partial_ih^k_j
+2\nabla^2h^k_0\partial_ih_{kj}\cr
&+2\nabla^2h^{00}\partial_ih_{0j}
-\partial_k\partial_lh^l_i\dot h^k_j
-\nabla^2h^k_i\dot h_{kj}\Bigr)\cr
+\lambda\Omega+\lambda_i\Omega^i
&+\rho\Theta+\rho_i\Theta^i
+\rho_{ij}\Theta^{ij}
+\xi\Lambda\,.&(3.10)\cr}$$

\bigskip
\noindent Using the consistency condition and making properly
combinations of the final constraints in order to obtain the biggest
number of first-class ones we get the final set

\medskip
({\it i}) \underbar {1st class contraints}

$$\eqalignno{\Omega&=\pi^{00}\approx0\,,&(3.11a)\cr
\Theta&=s^{00}\approx0\,,&(3.11b)\cr
\Theta^i&=s^{0i}\approx0\,,&(3.11c)\cr
\Gamma^i&=\partial_js^{ij}-\pi^{0i}
-{1\over2}\,\Bigl(\partial_jh^{ji}-\partial^ih^j_j\Bigr)
+{\kappa^2\over4\mu}\,\Bigl(
2\epsilon^{ji}\partial_k\partial_jh^{k0}\cr
&\phantom{=\partial_js^{ij}\,\,}
+\epsilon^{ij}\nabla^2h_{0j}
-\epsilon^{jk}\partial_k\dot h^i_j
+\epsilon^{ij}\partial_k\dot h^k_j\Bigr)
\approx0\,,&(3.11d)\cr
\xi^i&=\partial_j\pi^{ij}
+{1\over2}\Bigl(\nabla^2h^i_0
+\partial_j\partial^ih^{j0}\Bigr)
-{\kappa^2\over4\mu}\,\epsilon^{jk}\partial_j\,\Bigl(
\partial_l\partial^ih^l_k
+\nabla^2h^i_k\Bigr)\approx0\,,&(3.11e)\cr
\Gamma&=\partial_i\partial_js^{ij}
-{1\over2}\Bigl(\partial_i\partial_jh^{ij}
+\nabla^2h^i_i\Bigr)
+{\kappa^2\over2\mu}\epsilon^{ij}\partial_i\,
\Bigl(4\nabla^2h_{0j}+3\partial_k\dot h^k_j\Bigr)
\approx0\,.&(3.11f)\cr}$$

\bigskip
({\it ii}) \underbar {2nd class contraints}

$$\eqalignno{\Theta^{ij}&=s^{ij}
-\Bigl[{\kappa^2\over2\mu}\,\epsilon^{ik}\,
\Bigl(\partial_kh^j_0+\partial^jh_{0k}
-{1\over2}\,\dot h^j_k\Bigr)
+(i\,\leftrightarrow\,j)\Bigr]
\approx0\,,&(3.12a)\cr
\Lambda&=\pi^i_i+{1\over2}\,\dot h^i_i
-{\kappa^2\over2\mu}\,
\epsilon^{ij}\partial_j\partial_kh^k_i
\approx0\,.&(3.12b)\cr}$$

\bigskip
\noindent In fact we notice that there is just one degree of freedom
for D=3.
\medskip
We use the Dirac method iteratively. The elimination of the
second-class constraints above results in the following preliminar
brackets

$$\eqalignno{\bigl\{h^{ij}(x),\,\pi_{kl}(y)\bigr\}^\ast
&={1\over2}\,\bigl(\delta^i_k\,\delta^j_l
+\delta^i_l\,\delta^j_k\bigr)\,\delta^{(2)}(\vec x-\vec y\,)\,,\cr
\bigl\{\dot h^{ij}(x),\,s_{kl}(y)\bigr\}^\ast
&={1\over4}\,\bigl(\delta^i_k\,\delta^j_l+\delta^i_l\,\delta^j_k
-\eta^{ij}\,\eta_{kl}\bigr)\,\delta^{(2)}(\vec x-\vec y\,)\,,\cr
\bigl\{\dot h^{ij}(x),\,\dot h_{kl}(y)\bigr\}^\ast
&=-{\mu\over4\kappa^2}\,\bigl(
\epsilon^i\,_k\,\delta^j_l+\epsilon^i\,_l\,\delta^j_k
+\epsilon^j\,_k\,\delta^i_l+\epsilon^j\,_l\,\delta^i_k
\bigr)\,\delta^{(2)}(\vec x-\vec y\,)\,,\cr
\bigl\{s^{ij}(x),\,s_{kl}(y)\bigr\}^\ast
&=-{\kappa^2\over16\mu}\,\bigl(\epsilon^i\,_k\,\delta^j_l
+\epsilon^i\,_l\,\delta^j_k+\epsilon^j\,_k\,\delta^i_l
+\epsilon^j\,_l\,\delta^i_k\bigr)\,\delta^{(2)}(\vec x-\vec y)\,,\cr
\bigl\{h^{ij}(x),\,\dot h_{kl}(y)\bigr\}^\ast
&=-\eta^{ij}\,\eta_{kl}\,\delta^{(2)}(\vec x-\vec y\,)\,,\cr
\bigl\{\dot h^{ij}(x),\,\pi_{kl}(y)\bigr\}^\ast
&=-{\kappa^2\over4\mu}\,\eta^{ij}\,\bigl(
\epsilon^m\,_l\,\partial_m\partial_k
+\epsilon^m\,_k\,\partial_m\partial_l\bigr)\,
\delta^{(2)}(\vec x-\vec y\,)\,,\cr
\bigl\{h^{00}(x),\,\pi_{00}(y)\bigr\}^\ast
&=\delta^{(2)}(\vec x-\vec y\,)\,,\cr
\bigl\{\dot h^{00}(x),\,s_{00}(y)\bigr\}^\ast
&=\delta^{(2)}(\vec x-\vec y\,)\,,\cr
\bigl\{h^{0i}(x),\,\pi_{0j}(y)\bigr\}^\ast
&={1\over2}\,\delta^i_j\,\delta^{(2)}(\vec x-\vec y)\,.
&(3.13)\cr}$$

\bigskip
\noindent The remaining brackets are zero. To calculate the final
Dirac brackets we have to fix the gauge. Here we also choose the
corresponding of the Coulomb gauge (2.13), plus the further
conditions (2.14a) and (2.14b). Consistency with the equations of
motion permit us also to choose

$$\eqalignno{\dot h^{00}&\approx0\,,\cr
\dot h^{0i}&\approx0\,,\cr
\dot h^i_i&\approx0\,,\cr
\partial_ih^{ij}&\approx0\,.&(3.14)\cr}$$

\bigskip
\noindent With this gauge choice, the set of constraints given by
(3.11) becomes second-class. It is then possible to calculate the
Dirac brackets. After a hard algebraic work we get

$$\eqalignno{\bigl\{h^{ij}(x),\,\pi_{kl}(y)\bigr\}^D=
&-\Bigl[\eta^{ij}\eta_{kl}
+\eta^{ij}{\partial_k\partial_l\over\nabla^2}
+\eta_{kl}{\partial^i\partial^j\over\nabla^2}\cr
&-{1\over2}\bigl(\delta^i_k\delta^j_l
+\delta^i_k{\partial^j\partial_l\over\nabla^2}
+\delta^i_l{\partial^j\partial_k\over\nabla^2}
+(i\,\leftrightarrow\,j)\bigr)\Bigr]\,
\delta^{(2)}(\vec x-\vec y\,)\,,&(3.17a)\cr
\bigl\{\pi^{ij}(x),\,\pi_{kl}(y)\bigr\}^D=
&-{\kappa^2\over4\mu}\,\Bigl[
\bigl(\epsilon_{ml}\partial^m\partial_k
+\epsilon_{mk}\partial^m\partial_l\bigr)\,
\bigl(\eta^{ij}+2{\partial^i\partial^j\over\nabla^2}\bigr)\cr
&-2\bigl(\eta_{kl}+2{\partial_k\partial_l\over\nabla^2}\bigr)
\,\bigl(\epsilon^{mi}\partial_m\partial^j
+i\,\leftrightarrow\,j\bigr)\Bigr]\,
\delta^{(2)}(\vec x-\vec y\,)\,,&(3.17b)\cr
\bigl\{\dot h^{ij}(x),\,\dot h_{kl}(y)\bigr\}^D=
&-{\mu\over4\kappa^2}\,\Bigl(\epsilon^i\,_l\delta^j_k
+\epsilon^i\,_k\delta^j_l
+i\,\leftrightarrow\,j\Bigr)\,
\delta^{(2)}(\vec x-\vec y\,)\,,&(3.17c)\cr
\bigl\{\dot h^{ij}(x),\,s_{kl}(y)\bigr\}^D=
&{1\over4}\,\bigl(\delta^i_k\delta^j_l+\delta^i_l\delta^j_k
-\eta^{ij}\eta_{kl}\bigr)\,
\delta^{(2)}(\vec x-\vec y\,)\,,&(3.17d)\cr
\bigl\{s^{ij}(x),\,s_{kl}(y)\bigr\}^D=
&-{\kappa^2\over16\mu}\Bigl(\epsilon^i\,_l\delta^j_k
+\epsilon^i\,_k\delta^j_l
+i\,\leftrightarrow\,j\Bigr)\,
\delta^{(2)}(\vec x-\vec y\,)\,,&(3.17e)\cr
\bigl\{h^{ij}(x),\,\dot h_{kl}(y)\bigr\}^D=
&\Bigl(\eta^{ij}+{\partial^i\partial^j\over\nabla^2}\Bigr)\,
\Bigl(\eta_{kl}+2{\partial_k\partial_l\over\nabla^2}\Bigr)
\delta^{(2)}(\vec x-\vec y\,)\,,&(3.17f)\cr
\bigl\{\dot h^{ij}(x),\,\pi_{kl}(y)\bigr\}^D=
&{\kappa^2\over2\mu}\,\Bigl(
\epsilon^m\,_l\partial_m\partial_k
+l\leftrightarrow k\Bigr)\,
\Bigl(\eta^{ij}+{\partial^i\partial^j\over\nabla^2}\Bigr)\,
\delta^{(2)}(\vec x-\vec y\,)\,,&(3.17g)\cr
\bigl\{h^{ij}(x),\,s_{kl}(y)\bigr\}^D=
&{\kappa^2\over4\mu\nabla^2}\,\eta^{ij}\,\Bigl(
\epsilon^m\,_l\partial_m\partial_k
+l\leftrightarrow k\Bigr)\,
\Bigl(\eta^{ij}+{\partial^i\partial^j\over\nabla^2}\Bigr)\,
\delta^{(2)}(\vec x-\vec y\,)\,,&(3.17h)\cr
\bigl\{\pi^{ij}(x),\,s_{kl}(y)\bigr\}^D=
&{\kappa^4\over8\mu^2}\,\Bigl(\delta^i_k\partial^j\partial_l
+{\partial^i\partial^j\partial_k\partial_l
\over\nabla^2}+i\leftrightarrow j;\,
l\leftrightarrow k\Bigr)\,
\delta^{(2)}(\vec x-\vec y\,)\,,&(3.17i)\cr
\bigl\{\dot h^{ij}(x),\,\pi_{0k}(y)\bigr\}^D=
&\Bigl({1\over2}\eta^{ij}+{\partial^i\partial^j\over\nabla^2}
\Bigr)\,\partial_k
\delta^{(2)}(\vec x-\vec y\,)\,,&(3.17j)\cr
\bigl\{s^{ij}(x),\,\pi_{0k}(y)\bigr\}^D=
&{\kappa^2\over4\mu}\,
\Bigl(\epsilon^{mi}\partial_m\partial^j
+\epsilon^{mj}\partial_m\partial^i\Bigr)\partial_k\,
\delta^{(2)}(\vec x-\vec y\,)\,.&(3.17k)\cr}$$

\bigskip
\noindent One can check the validity of the brackets above by showing
that they are strongly consistent with all the constraints.

\bigskip
We notice that there are no problem with ordering operators. So, the
above brackets can be transform without problems to commutators by
means of the usual rule of quantization given by (2.18). The Feynman
propagators among the fields $h^{ij}$ are (details of calculating
Feynman propagators when there are higher derivatives can be found in
references~[8])

$$\eqalignno{<0\vert T\bigl(h^{ij}(x)h_{kl}(y)\vert0>
=&i\int{d^3k\over(2\pi)^3}\,{{\rm e}^{ik(x-y)}
\over k^2\bigl({\strut\mu^2\over\displaystyle\kappa^4}
-4k^2_0\bigr)}\,
\biggl[4\kappa^2k_0^2{k_kk_l\over\vec k^2}\,\eta^{ij}\cr
&-{\mu^2\over\kappa^4}\,\Bigl(\eta^{ij}\eta_{kl}
+{k^ik^j\over\vec k^2}\,\eta_{kl}
+2{k_kk_l\over\vec k^2}\,\eta^{ij}
+2{k^ik^jk_lk_k\over\vec k^4}\Bigr)\cr
&-{\mu^2\over2\kappa^2}\,\Bigl(\eta^j_l\eta^i_k
+\eta^j_k\eta^i_l-2\eta_{kl}\eta^{ij}\Bigr)\cr
&+8k_0^2\,\Bigl({1\over2}\eta_{ij}\eta_{kl}
+{k_kk_l\over\vec k^2}\,\eta_{ij}
+{1\over2}\,{k_ik_j\over\vec k^2}\,\eta_{kl}
+{k_ik_jk_kk_l\over\vec k^4}\Bigr)\cr
&-{\mu^2\over\kappa^2\vec k^2}\Bigl(k_kk_l\eta^{ij}+k^ik^j\eta_{kl}
-\bigl(k^ik_l\eta^j_k+i\leftrightarrow j; k\leftrightarrow l
\bigr)\Bigr)\cr
&+{i\mu k_0\over\vec k^2}\,
\Bigl({1\over2}\epsilon^{pi}k_pk^j\eta_{kl}
+\epsilon^i_kk^jk_l
+i\leftrightarrow j; k\leftrightarrow l\Bigr)\cr
&-{i\mu k_0\over2}\Bigl(\epsilon^k\,_i\eta^j_l
+i\leftrightarrow j; k\leftrightarrow l\Bigr)\biggr]\,.&(3.18)\cr}$$

\bigskip
\noindent We observe that there is a massive (topological) pole given
by $\mu/2\kappa^2$. This is in agreement with previous results found
in literature~[3,4,9] and show the consistency of the quantization procedure
we have used.

\vfill\eject
{\titlea IV. Conclusion}
\bigskip
We have studied the quantization of the Einstein gravitational theory
in the formalism of weak gravitational fields. This is a high
constrained system and we have used the Hamiltonian Dirac method to
quantize it. We have given a particular emphasis to the case of the
spacetime dimension D=2+1 where a CS term can be added. The final
propagators in this case exhibit a massive (topological) pole.  This
result is in agreement with previous one found in literature that use
other quantization procedure.

\vskip 1cm
{\titlea Acknowledgment}
\bigskip
This work was supported in part by Conselho Nacional de
Desenvolvimento Cient\\fi- co e Tecnol\'ogico - CNPq (Brazilian
Research Agency).

\vfill\eject
{\titlea References}
\bigskip
\item {[1]} See, for example, N.D. Birrel and P.C.W. Davies: {\it
Quantum fields in curved space} (Cambridge University Press,
Cambridge, 1984) and references therein.
\item {[2]} See for example, M.B. Green, J.H. Schwarz and E.Witten,
{\it Superstring theory} (Cambridge University Press, Cambridge, 1984)
and references therein.
\item {[3]} S.Deser, R. Jackiw and S. Templeton, Phys. Rev. Lett. 48
(1982) 975; Ann. Phys. (NY) 140 (1982) 372.
\item {[4]} S. Deser and Z. Yang, Class. Quantum Grav.7 (1990) 1603;
\item {[5]} P.A.M. Dirac, Can. J. Math. 2 (1950) 129; {\it Lectures
on quantum mechanics} (Yeshiva University, New York, 1964); See also
A. Hanson, T. Regge and C.  Teitelboim, {\it Constrained Hamiltonian
systems} (Academia Nazionale dei Lincei, Rome, 1976); K.
Sundermeyer, {\it Constrained Dynamics}, Lectures Notes on Physics
(Springer, New York, 1982), Vol. 169.
\item {[6]} L. Castellani, Ann. Phys. (N.Y.) 143 (1982) 357.
\item {[7]} For recent references on canonical quantization of higher
derivative systems we mention V.V. Nesterenko, J. Phys. A22 (1989)
1673; C. Battle, J. Gomis, J.M. Pons and N. Rom\'ana-Roy, J. Phys.
A21 (1988) 2693; C.A.P. Galv\ao and N.A. Lemos, J. Math. Phys. 29
(1988) 1588; C.G. Bollini and J.J. Giambiagi, Phys. Rev. D39 (1989)
1169; J. Barcelos-Neto and N.R.F. Braga, Acta Phys. Polonica B20
(1989) 205.
\item {[8]} J. Barcelos-Neto and N.R.F. Braga, Mod. Phys. Lett. A4
(1989) 2195; J. Barcelos-Neto and C.P. Natividade, Z. Phys. C51
(1991) 313.
\item {[9]} C. Pinheiro and G.O. Pires, {\it Extending the
Barnes-Rivers operators to D=3 topological gravity}, Preprint
CBPF/1993.

\bye